# Vacuum Phase Transitions and the Temperature of the Universe

V.G. Lapchinsky, V.A Rubakov  and A.V. Veryaskin*

*University of Western Australia, School of Physics, Mathematics and Computing
 35 Stirling Highway, Crawley WA 6009, Australia
 Email: alexey.veryaskin@uwa.edu.au

**Abstract**

The authors discuss a possibility that the present great value of $aT$ ( $a$ is the radius of spatial curvature and $T$ is the temperature of the Universe ) was generated by first order vacuum phase transitions. In Coleman-Weinberg type models such transitions are natural and they rise the value of $aT$ from unity to $\exp(\text{const}/a_g)$. Some GUTs are briefly discussed in this contest.

**Introduction**

The manuscript presented below has never been published elsewhere except for a preprint of Institute for Nuclear Research (INR) of Academy of Sciences of USSR published in 1982 [1]. It was a result of an informal collaboration between Vladimir Lapchinsky (INR), Valery Rubakov (INR) and Alexey Veryaskin ( Sternberg State Astronomical Institute of Moscow State University ) which lasted from 1974 for about a decade. The preprint has not been properly mentioned or cited in relevant publications except for a few self-citations by Rubakov [2] and Lapchinsky et al [3]. It was first cited by Andrei Linde in his centenary paper on the new cosmic inflation theory published in 1982 [4]. Recently, Linde mentioned it again in his presentation at the Rubakov's memorial conference held in Yerevan in 2023 [5]. Interestingly, a record about it has been found in the Steven Weinberg personal archive held at the Harry Ransom Center at the University of Texas at Austin [6]. The preprint's content may only be of interest as a historical example of a fundamental physics trial and error research conducted by some of my generation of physicists trying to understand the Universe we all live in. I am grateful to Mikhail Shaposhnikov for our recent discussions about this preprint and its content who pointed out a relevant work by Witten [7] which was not known to the authors of the preprint at the time of composing it in 1981[#].

Alexey (Alex) Veryaskin
Auckland, New Zealand
July 2025

[#]The preprint's content has not been modified and is presented below in its original form as published by the INR in 1982 (see page 7).

**Introduction References**

It is widely accepted that our Universe is extremely hot [1]. It may by characterized by at least two dimensionless quantities, namely i) $(n_B - n_{\overline{B}})/n_\gamma$, where $n_B$, $\overline{n}_B$, $n_\gamma$ are the baryon, antibaryon and photon densities, and ii) $(k_B/\hbar c)aT$, where $a$ is the conformal factor, i.e. radius of spatial curvature of the Universe* and $T$ is the temperature of the Universe. The first quantity characterizes the baryon asymmetry of the Universe and its relevance to particle physics has been recently recognized [2, 3]. The second quantity may be viewed roughly as a measure of the total entropy inside the sphere of radius $a$ [1]. It is time independent in the case of adiabatic expansion and an order of magnitude estimate of its present value is

$$(aT)_{present} \sim (1/H_O)\, 3K \sim 10^{28} - 10^{29} \tag{1}$$

( hereafter $\hbar = c = k_B = 1$ ), where $H_O = 10^{10}$ (years)$^{-1}$ is the present value of the Hubble constant.

Such a great value of (1) is striking ( note, that in the standard hot big-bang model it is an arbitrary parameter ). It would be attractive to assume that "in the very beginning" all the dimensionless characteristics of the Universe were of order of unity ( say, at $a = l_{pl} = 10^{-33}$ cm, energy density, entropy density, pressure etc. were Plank ones ). If one adopts this philosophy, the present value of aT should be calculable and there should exist an effective mechanism of heating the Universe in intermediate stages. This mechanism can be provided by vacuum phase transitions**. It has been shown recently [5] that in cold charge-asymmetric Universe the first order phase transition is possible and under certain assumptions concerning the microscopic parameters ( coupling constants, masses etc. ) as well as macroscopic ones, the present value of $(n_B - n_{\overline{B}})/n_\gamma$, can be generated ( the second quantity, aT, has not been considered in [5] ). In the present paper we adopt a completely different point of view. We consider initially charge-symmetric Universe assuming the generation of the present baryon asymmetry during the process of evolution [2, 3].

---

*We do not consider here the possibility that the energy density in the Universe exactly equals to critical one. In this case the Universe is spatially flat and the scale factor has no physical meaning.

**We do not consider the case of the Universe with the critical density. In this case the Universe is spatially flat and the scale factor has no physical meaning.

Our aim is to show that in the models with spontaneous symmetry breaking due to quantum corrections ( Coleman-Weinberg mechanism [6] ), first order phase transitions are natural and they raise the value of aT from unity to exponentially great numbers ( aT ≈ exp ( const/$a_g$ ) ). We shall see, however, that in Weinberg-Salam model as well as in SU(5) grand unified theory*** [7] this mechanism leads to the values of aT much smaller than (1). We shall argue that GUTs with several mass scales (and presumably with early unification) are preferable from this point of view ( the examples are [SU (4)]$^4$ and [SU(8)]$^2$ GUTs [9] ).

As a toy model we consider SU(N) gauge model with one fundamental Higgs multiplet. We assume that the Universe is open and its expansion rate is small at least before the phase transition ( the validity of the latter assumption will be discussed in the end of the paper ) . Our basic assumption is the Coleman-Weinberg mechanism of symmetry breaking , i. e. the absence of mass parameter in the tree effective potential at a = ∞ , T = 0. We shall not consider the effects of fermions on the effective potential, assuming small Yukawa couplings. Then the vacuum expectation value $\phi_0$, of the scalar field at T = 0 is a solution of the equation [6, 10]

$$\bar{\lambda}(\phi) = 0(\bar{g}^4(\phi_0)) \qquad (2)$$

where $\bar{\lambda}$ and $\bar{g}$ are the running scalar self-coupling and running gauge coupling respectively. Hereafter we assume $\phi_0 \ll m_{pl} = 10^{19}$ GeV. The relevant property of Coleman-Weinberg type models is that the effective potential at nonzero temperature [11] has a minimum ( at least local ) at $\phi = 0$, this minimum being unique at T >> $M_v$ where $M_v = g\phi_0$ is the mass of vector bosons ( up to inessential numerical factors ). Thus, under the above assumptions the symmetry is restored "in the very beginning". When the temperature becomes smaller than $M_v$, the second minimum of the effective potential appears at nonzero $\phi$ and at at T << $M_v$ the effective potential takes the form shown in Fig. 1 with $\Delta \phi \approx T/g$ . The phase transition becomes possible, but the probability of the phase transition is exponentially small up to exponentially small temperatures. The last statement is best illustrated by the model of one massless scalar field with $(+\lambda/4!)\phi^4)$ self-interaction ( negative sign of the potential ).

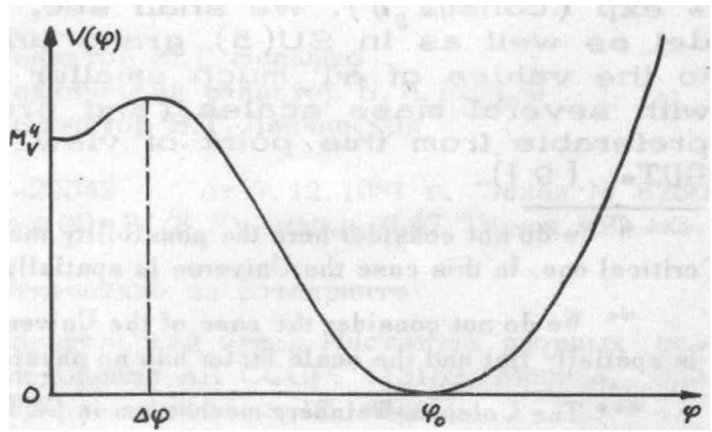

Fig. 1. The effective potential of the toy model at small temperature (the original preprint's copy).

***The Coleman-Weinberg mechanism in GUTs has been considered in [8].

The state with $<\phi> = 0$ is unstable in this model but the probability of the fluctuation with the characteristic value of the field ρ is proportional to exp(-8$π^2$/$\bar{\lambda}$(ρ)), as can be extracted from [12, 13]. Just the same is true for our toy- model, but the value of ρ is bounded from below by $\Delta \phi \approx$ T/g, so that the decay probability of the state with $<\phi> = 0$ is suppressed by exp(-8$π^2$/$\bar{\lambda}$(T). We find that the phase transition is impossible up to exponentially small temperatures, but when $\bar{\lambda}$(T) becomes of order of unity, the probability of the phase transition becomes large and vacuum intensively "boils".

We obtain an equation for the temperature of the phase transition $T_{crit}$ ( and through aT ~ 1 for the scale factor $a_{crit}$ at the moment of the phase transition)

$$\bar{\lambda}(T_{crit}) = \bar{\lambda}(a_{crit}^{-1}) = 1 \tag{3}$$

As a result of the phase transition the energy density ~ M is liberated and after the phase transition the value of aT becomes equal to

$$aT = a_{crit} M_v \tag{4}$$

This value of aT remains constant up to the present stage of the evolution of the Universe ( or up to the next phase transition, see below ).

The solution of Gell-Mann-Low equations with the initial conditions (2) is [14]

$$\bar{\lambda}(\rho) = \bar{g}^2(\rho) F[\bar{g}^2(M_v)/\bar{g}^2(\rho)]$$

where the function F depends also on N and $b_0$ = ( 11/3) N -1/6 - (fermionic contributions), its explicit form can be found, e.g., in [10]. From (3) and (4) we get

$$aT = \exp\left[\frac{2\pi}{b_0 a_g}(1 - \exp(-C))\right], \quad a_g = \frac{\bar{g}^2(M_v)}{4\pi} \tag{5}$$

where C = C (N, $b_0$) is a solution of F (c ) = ∞ . Eq.(5) is a crude estimate; we believe, however, that it is a good starting point for the discussion of realistic models. For SU(5) group $a_g$ = 1/45 and realistic fermionic content ( three generations of 5 + 10 ) [7], Eq.( 5) gives

$$(aT)^{(3)}_{SU(5)} = 10^9 \tag{6}$$

Moreover, irrespectively of the number of fermionic generations, the upper bound is

$$(aT)^{(any)}_{SU(5)} < 10^{18} \tag{7}$$

The realistic SU(5)-GUT is hard to analyse because of the complexity of Higgs sector, but one can show, that ( 6) is an upper bound for three generations and 5 + 24 Higgs scalars. We also expect ( 7) to be unaltered by the inclusion of Higgs 24-plet. The direct application of (5) to Weinberg-Salam model with three generations gives aT ~ $10^{20}$.

However, the arguments that have led to (5) are incorrect in this case because of strong contributions to the effective potential at small T ( for example, the linear term ($<qq>M_q/\phi_0$)$\phi$ is generated by strong interactions [15] ) and we expect $T_{crit} \sim \Lambda_c = 100 \div 500$ MeV and

$$(aT)_{WS} \sim 10^3 \qquad (8)$$

Thus, both in SU(5) theory and in Weinberg-Salam one the calculated value of aT is much smaller than (1). The situation is better in the theories with several mass scales. One can show, that for the gauge hierarchy

$$G_1 \to G_2 \to G_3 \to \cdots$$

$$M_v^{(1)} \to M_v^{(2)} \to M_v^{(3)} \to \cdots$$

the corresponding phase transitions in the Universe under the above assumptions lead to the resulting value of aT equal to

$$(aT) = (aT)_1 (aT)_2 (aT)_3 \ldots$$

where $(aT)_k$ is the value of aT which would be generated by k-th phase transition $G_k \to G_{k-1}$, if the initial value of am was of order of unity. Thus GUTs with several mass scales are preferable from the point of view advocated in this paper.

A few comments are in order:

1) In the above discussion we completely ignored the corrections to the effective potential arising from the non-Euclidean character of space [16, 17]. For aT ~ 1 these are comparable. with the temperature ones . Moreover, one can adopt the extremal point of view and assume aT = 0 before the phase transition ( cold Universe ). If the scalar - gravity coupling is minimal ( no conformal term $RSp\phi^2$ in the Lagrangian ), the one-loop effective potential in our toy model at small $\phi$ is[#]

$$V(\phi) = -\frac{N^2-1}{N} \frac{g^2 \phi^2}{8\pi^2 a^2} \ln\frac{\phi}{\phi_0} + O(g^4 \phi^4 \ln \frac{\phi}{\phi_0})$$

and at small $a^{-1}$ it has just the same form as shown in Fig.1 but with $\Delta\phi = 1/a_g$. The above discussion is directly applicable to this case and (5) remains true.

2) The above assumption concerning the low rate of the expansion of the Universe is not generally valid. This is because of the existence of $\Lambda$-term ( see [5] and references therein)

$$\Lambda = (m_{pl})^{-2} M_v^4$$

before the phase transition. One can show that the expansion rate is indeed low if

$$a_{crit}^{-1} \gg M(M_v/m_{pl}), \qquad (9)$$

---

[#] The evaluation of the one-loop effective potential in the background open static Friedman-Robertson-Walker metric follows Ref.[17] and will be published elsewhere.

otherwise our main arguments are incorrect ( e.g., the very notion of the effective potential is obscure). As has been shown by Linde, if (9) does not hold, the phase transition is really impossible. This remark along with Eqs.(6)-(8) shows that it is quite unlikely for the experimental value (1) to be developed through the discussed mechanism in GUTs with all mass scales near $m_{pl}$ ( except $M_w$ ) such as SU(5) or SO(10). One can speculate that this is an argument in favour of early unification. Whether the present mechanism of aT-generation is compatible with the popular mechanisms of the baryon asymmetry generation [2,3] remains unclear. We hope to turn to this question in future.

We are deeply indebted to A. Yu. Ignatiev, N. V. Krasnikov, V. A. Kuzmin, A.D. Linde, V. A. Matveev, V. I. Nekrasov, M. E. Shaposhnikov and I. I. Tkachov for many helpful discussions and comments.

After this work has been completed, the paper by A. Guth (Phys. Rev. D23 (1981) 347) appeared, in which the problem of a large value of aT, as well as the possibility of the generation of this value in the process of the first order phase transition, were considered. Unlike A. Guth, we study here in a detail the Coleman-Weinberg case of symmetry breaking and arrive to a quite different picture of the phase transition, which is characterized by a large nucleation rate below the critical temperature $T_{crit}$.

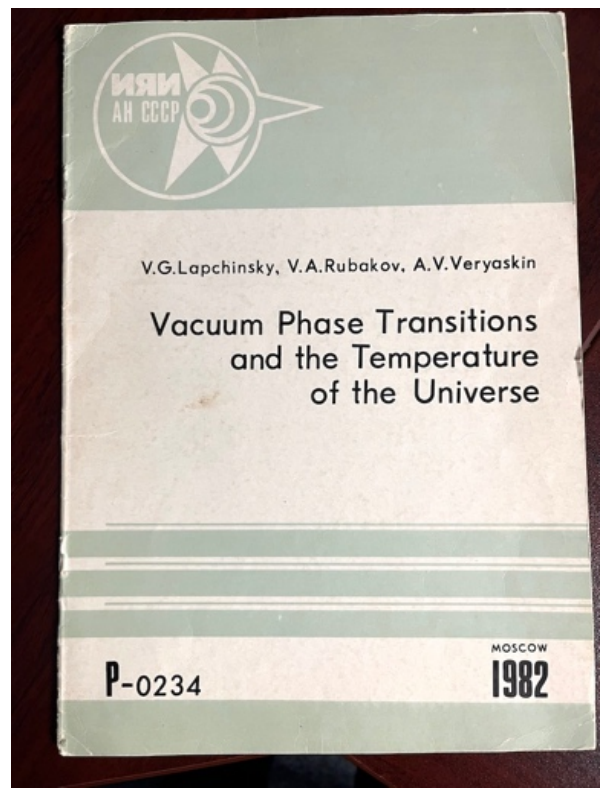